\documentclass[a4paper,twocolumn, nofootinbib,superscriptaddress,aps,pra, 10pt,eqsecnum,notitlepage,showkeys,showpacs]{revtex4-1}
\usepackage{multirow}
\usepackage{gensymb}
\usepackage{amsmath}
\usepackage{dcolumn}    
\usepackage{ifpdf}
\usepackage{amssymb,lineno,amsfonts}
\usepackage{graphicx}   
\usepackage{bm}         
\usepackage{bbm}
\usepackage{mathrsfs}
\usepackage{upgreek}
\usepackage{float}
\usepackage{mathtools}
\usepackage{epstopdf}
\usepackage{setspace}
\usepackage{hyperref}
\usepackage{natbib}
\usepackage{esvect}
\usepackage[usenames,dvipsnames]{xcolor}
\definecolor{med-blue}{RGB}{25,25,112}
\hypersetup{colorlinks, linkcolor={blue},citecolor={blue}, urlcolor={blue}}
\usepackage{times }
\usepackage [english]{babel}
\usepackage{siunitx}
\usepackage{multirow}
\usepackage [autostyle, english = american]{csquotes}
\MakeOuterQuote{"}

\begin{document}
\title{Frustration, strain and phase co-existence in the mixed valent hexagonal iridate Ba$_{3}$NaIr$_{2}$O$_{9}$}
\author{Charu Garg}
\affiliation{Department of Physics, Indian Institute of Science Education and Research, Dr. Homi Bhabha Road, Pune 411008, India}
\author{Antonio Cervellino}
\affiliation{Swiss Light Source, Paul Scherrer Institute, CH-5232 Villigen, Switzerland}
\author{Sunil Nair}
\affiliation{Department of Physics, Indian Institute of Science Education and Research, Dr. Homi Bhabha Road, Pune 411008, India}
\affiliation{Centre for Energy Science, Indian Institute of Science Education and Research, Dr. Homi Bhabha Road, Pune 411008, India}
\date{\today}

\begin{abstract} 
Using detailed synchrotron diffraction, magnetization, thermodynamic and transport measurements, we investigate the relationship between the mixed valence of Ir, lattice strain and the resultant structural and magnetic ground states in the geometrically frustrated triple perovskite iridate Ba$_{3}$NaIr$_{2}$O$_{9}$. We observe a complex interplay between lattice strain and structural phase co-existence, which is in sharp contrast to what is typically observed in this family of compounds. The low temperature magnetic ground state is characterized by the absence of long range order, and points towards the condensation of a cluster glass state from an extended regime of short range magnetic correlations.
\end{abstract} 

\maketitle
\section{Introduction}
Geometrically frustrated magnets- where triangular lattice antiferromagnets (TLAFs) are considered to be an archetype- remain at the forefront of contemporary condensed matter \cite{doi:10.1146/annurev-conmatphys-020911-125138,doi:10.1146/annurev.ms.24.080194.002321,doi:10.1063/1.2186278}. Of particular interest in the recent years have been a number of Ruthenium and Iridium based perovskite variants which stabilize in an inherently frustrated environment. In general, the stabilization of a particular structural ground state depends on the tolerance limit of the corresponding symmetry which in turn is related to the relative ionic radii of the constituent elements. For instance, in the perovskite ABO$_{3}$, introducing a bigger element at the $A$ and $B$ sites can progressively tune the lattice from a high symmetry hexagonal to a lower symmetry orthorhombic, or even a monoclinic one \cite{doi:https://doi.org/10.1002/9780470022184.hmm411}. The same is true for the double (A$_{2}$BB$^{'}$O$_{6}$) and triple layered perovskites (A$_{3}$BB$^{'}_{2}O_{9}$) as well, where it has been shown that $B$ site cations with higher atomic radii stabilizes in a lower symmetry \cite{ZHAO2009327,PhysRevB.97.064408,VASALA20151}. 

A relatively recent addition to this family of geometrically frustrated magnets are the Barium based triple perovskite iridates of the form Ba$_{3}$MIr$_{2}$O$_{9}$ (\textit{M}=alkali metal, alkaline earth metal, 3$d$ transition metal or lanthanides). The choice of the M-site cation strongly determines the crystallographic symmetry, which in turn inordinately influences the magnetic ground states. For example, in M=Zn${^{2+}}$, a close realization of the elusive J=0 state is observed whereas for M= Mg${^{2+}}$, Sr${^{2+}}$ and Ca${^{2+}}$, deviation from the non magnetic state in the form of antiferromagnetic exchange interactions, ferromagnetic and weak dimer like features are observed respectively \cite{PhysRevB.97.064408}. Another addition to this family is the newly reported Ba$_{3}$CoIr$_{2}$O$_{9}$, where Co${^{2+}}$, being a magnetic ion strongly influences exchange paths leading to weak ferromagnetism at low temperature and the highest magneto-structural transition temperature reported in the triple perovskite iridates \cite{garg2020evolution}. On the other hand,  Ba$_{3}$BiIr$_{2}$O$_{9}$ has been reported to exhibit a giant magneto-elastic transition accompanied by the opening of a spin gap \cite{BaBi}. The structure-property relationships in these systems are clearly driven by a complex interplay between the relative strengths of competing spin-orbit coupling (SOC), electronic correlation (U), and a hybridization interaction controlled by the Ir-O-Ir bond angle. Thus small perturbations, such as changes in lattice parameters caused by variations of different M ions, can tip the balance between the competing energies and ground states. 

In all the reported 6H hexagonal triple perovskite iridates, the ionic radii of the $M$ site cation lies in the range of 0.605$\AA$-0.947$\AA$, beyond which the internal pressure forces the lattice to stabilize in a lower symmetry. For example, the Ba$_{3}$CaIr$_{2}$O$_{9}$ system has been reported to stabilize in $C2/c$ monoclinic symmetry \cite{ZHAO2009327} which is in line with the expected structural ground state based on the tolerance limit. Interestingly, an exception appears to be the Ba$_{3}$NaIr$_{2}$O$_{9}$ system, which - in spite of the similar ionic radii of Na (1.02$\AA$) and Ca (1.00$\AA$) - has been reported to stabilize in the high symmetry hexagonal structure at room temperatures. In this report, we discuss this relatively un-investigated Na based triple perovskite iridate, where iridium is forced to be in the unconventionally high charge state of 5.5. We investigate polycrystalline specimens of this system using a combination of high resolution synchrotron diffraction, magnetization, resistivity and specific heat measurements. We observe that the lattice appears to accommodate strain as the temperature is reduced, which in turn precludes the stabilization of a lower symmetry structural phase. This is in contrast to what is typically observed in this class of materials. On the other hand, a very gradual and incomplete transformation to a low symmetry orthorhombic phase is observed, and the high symmetry hexagonal phase survives till the lowest measured temperatures. Measurements of the magnetization and specific heat point towards the existence of a extended cooperative paramagnetic regime characterized by short range magnetic correlations, which condenses into a cluster glass like state at low temperatures. 

\section{Experimental Details}
Polycrystalline specimens of Ba$_{3}$NaIr$_{2}$O$_{9}$ were synthesized by using the standard solid state reaction route. Stoichiometric amounts of high purity BaCO$_{3}$, Na$_{2}$O$_{3}$ and IrO$_{2}$ were thoroughly ground and then sintered at 1100$^{0}$C under oxygen atmosphere to maintain the high oxidation state of Iridium. The phase purity was confirmed by x-ray diffraction using a Bruker D8 Advance diffractometer with a Cu K$\alpha$ radiation. High resolution synchrotron x-ray diffraction data was collected using the Materials Science (MS) X04SA beam line (wavelength 0.56526$\lambda$) at the Swiss Light Source (SLS, PSI Switzerland). The finely ground powder was loaded in a glass capillary of diameter 0.3mm and was spun during the data acquisition at various temperatures between 5\,K and 300\,K. The structure was analyzed by Rietveld refinement using the FULLPROF suite \cite{Fullprof, doi:10.1107/S0021889869006558}. The structures shown in the manuscript are drawn using Vesta \cite{vesta}. The homogeneity and stoichiometry of the compound were also reconfirmed by energy dispersive x-ray (EDAX) from ZEISS Ultra Plus. Magnetization and physical property measurements were performed using a Quantum Design (MPMS-XL) SQUID magnetometer and a Physical Property Measurement System (PPMS) respectively. 

\section{Results and Discussion}
\begin{figure}
	\centering
	\includegraphics[scale=0.31]{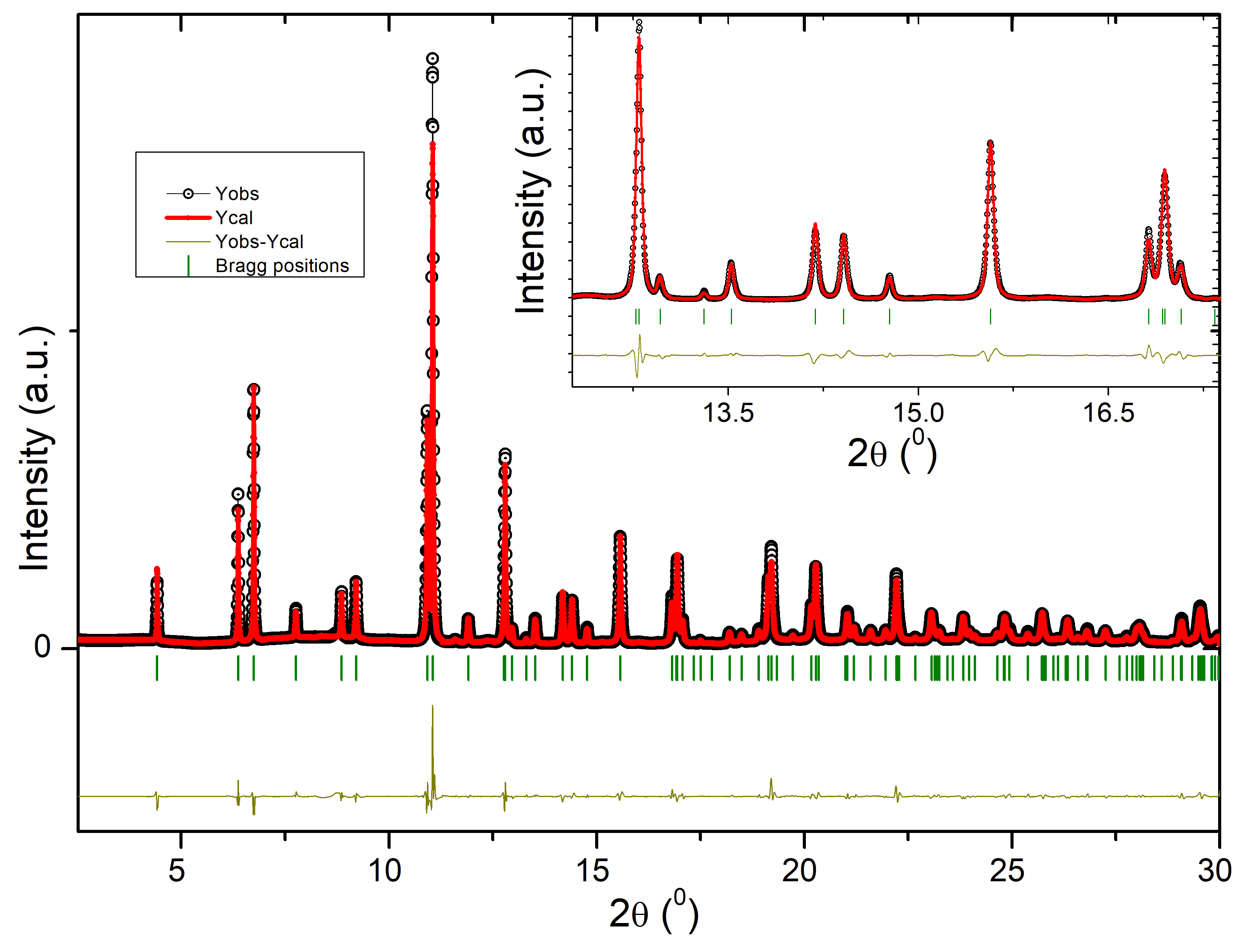}
	\caption{ Main panel: Fit to the Rietveld refinement of the synchrotron data at 295\,K for Ba$_{3}$NaIr$_{2}$O$_{9}$. The compound crystallizes in a 6H-hexagonal perovskite with space group P6$_{3}$/mmc (194). The calculated and the observed diffraction profiles are shown in red and black respectively. The vertical green lines indicates the Bragg positions and the brown line at the bottom is the difference between observed and calculated intensities. Inset: Enlarged view of the higher angle peaks and the corresponding fit.}
	\label{51}
\end{figure}
\begin{figure}
	\centering
	\includegraphics[scale=0.37]{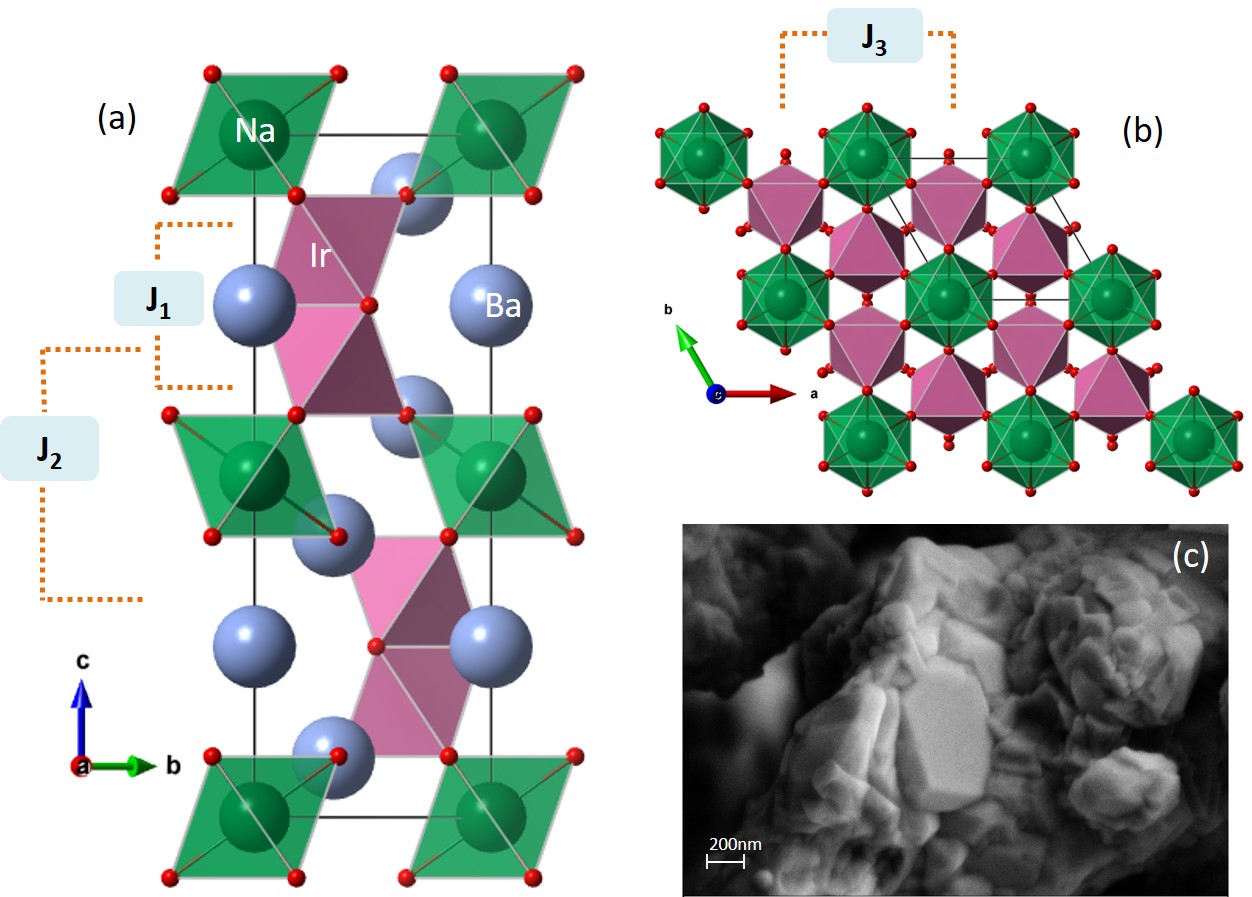}
	\caption{(a) A schematic representation of the crystal structure of  Ba$_{3}$NaIr$_{2}$O$_{9}$ using Vesta. Here pink and green octahedra represents Iridium and Sodium respectively and the Barium atoms are represented in Blue. (b) The projection of the structure along the c-axis is shown. The Iridium octahedra form a hexagonal ring surrounded by Sodium. (c) Scanning electron micrograph of the compound showing hexagonal facets.}
	\label{52}
\end{figure}
A Rietveld fit to the synchrotron diffraction data obtained at 300\,K is shown in  Fig.~\ref{51} where Ba$_{3}$NaIr$_{2}$O$_{9}$ is seen to stabilize in the high symmetry hexagonal ($P6_{3}/mmc$) symmetry, and the lattice parameters are deduced to be a = b =  5.86282(3)$\AA$, c = 14.61922(10)$\AA$ and $\alpha= \beta$ = 90$^{\circ}$; $\gamma$ = 120$^{\circ}$. This is in good agreement with previous reports\cite{LIGHTFOOT1990174,RIJSSENBEEK199965,DOI2001113,doi:10.1021/ic051344z}. The room temperature structure is illustrated in Fig.~\ref{52}(a) where face sharing octahedra (in pink) forms a Ir$_{2}$O$_{9}$ dimer and are connected via corners to NaO$_{6}$ octahedra (in green). Fig.~\ref{52}(b) represents the projection along the crystallographic $c$-axis where IrO$_{6}$ octahedra forms a hexagonal ring around the NaO$_{6}$ octahedra.  Since Na is in the +1 oxidation state, Ir is forced to stabilize in an atypical high oxidation state of +5.5. EDAX measurements were also used to confirm the stoichiometry. Since it is difficult to quantify the lighter elements (Na and O) using this technique, the atomic percentage ratio between heavy elements Ba and Ir was compared. The Ba:Ir ratio obtained from EDAX was observed to be 1.54 which is very close to the stoichiometric ratio of 3:2=1.5 expected from the chemical formula.  A scanning electron micrograph image is shown in Fig.~\ref{52}(c) where hexagonal facets - a reflection of the underlying  crystallographic symmetry - can be clearly seen.\\
\begin{figure}
	\centering
	\includegraphics[scale=0.8]{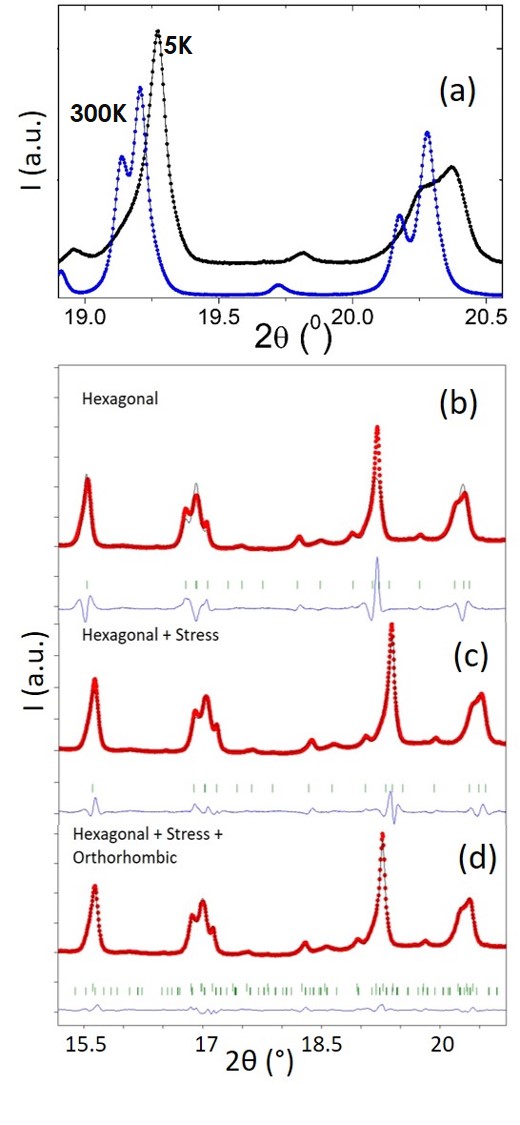}
	\caption{(a) Temperature evolution of synchrotron peaks at 5\,K (black) and 300\,K (blue). The lattice strain manifests in the form of broadening of diffraction peaks as evident by the highly anisotropic peak profile at 5K. (b,c,d) Attempts to fit the synchrotron diffraction data at 5\,K using various refinement models as indicated. }
	\label{53}
\end{figure}
\\A comparison of the temperature dependence of a few representative x-ray diffraction peaks as measured at the extreme temperatures of 5\,K and 300\,K is shown in Fig.~\ref{53}(a). As the temperature is lowered, the diffraction peaks shift to higher angles and also becomes anisotropic. The modification of the peak profile could either signal the presence of strain in the lattice or a transformation to a lower symmetry phase. The former could be a consequence of the large ionic radii which Na possesses, whereas the latter has been reported in a number of triple perovskite iridates earlier.  Since there were no additional peaks visible in the low temperature scan, the data was initially fit using a hexagonal model alone. These attempts were not successful, as is shown in the Fig.~\ref{53}(b). Addition of strain using the broadening model available in FullProf made the fit better as can be seen in Fig.~\ref{53}(c). This method is based on Stephens model \cite{Stephens:hn0085} of anisotropic broadening, where the refinement of microstrain covariance parameters S$_{400}$, S$_{004}$ and S$_{112}$ corresponds to strain along the 100, 001 and 101 hkl planes. Though strain does appear to have an impact on the low temperature phase, the fitting was still not satisfactory enough, which hints at the possible presence of an additional low symmetry phase at low temperatures. \\
\begin{figure}
	\centering
	\includegraphics[scale=0.45]{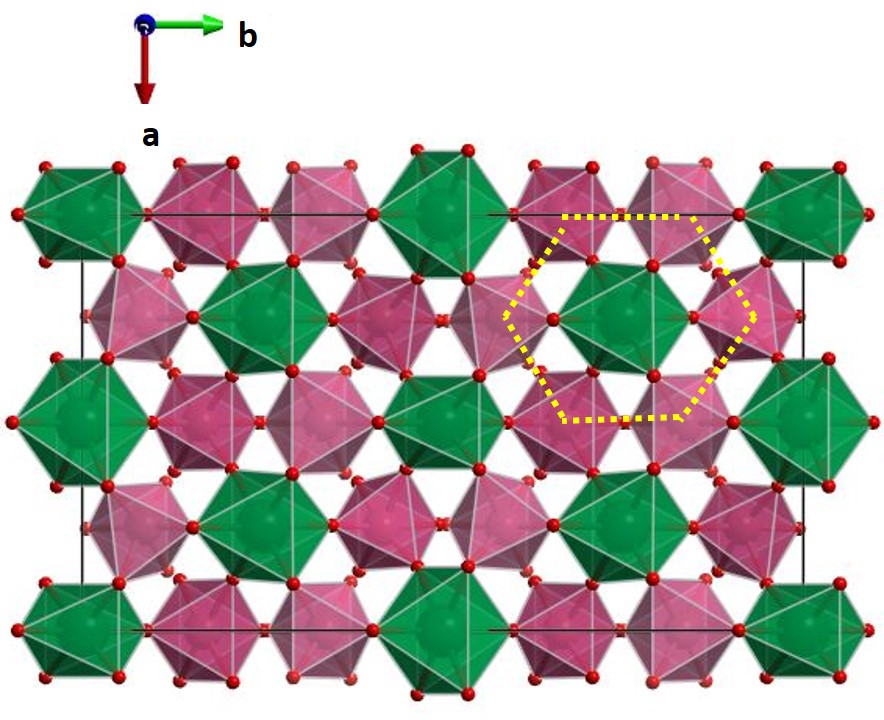}
	\caption{A schematic representation of the crystal structure of Ba$_{3}$NaIr$_{2}$O$_{9}$ using Vesta for the orthorhombic phase. Here pink and green octahedra represents Iridium and Sodium respectively. The Barium atoms are not shown for clarity. The yellow dotted line shows the hexagonal arrangement for Iridium octahedra. }
	\label{55}
\end{figure}
To identify the possible symmetry of the additional low temperature phase, existing literature in the ruthenate triple perovskite family was referred to, where multiple scenarios ranging from monoclinic ($P2/c$, $C2/c$) to orthorhombic ($Cmcm$), or even different structural models for the same compounds \cite{PhysRevLett.108.217205,doi:10.1021/ja0271781} have been reported. After exploring all these possible options, the orthorhombic (space group-Cmcm (63)) phase \cite{doi:10.1021/ja0271781} resulted in the best fit, with R$_{wp}$ and R$_{p}$ values of 3.24 and 2.47 respectively. The generated pattern was seen to match well with the high resolution synchrotron data as shown in Fig.~\ref{53}(d). The lattice parameters obtained from the fit for the additional orthorhombic phase at 5K are a= 11.6574(11)$\AA$, b=20.1975(21)$\AA$, c=14.5773(03)$\AA$ and $\alpha=\beta=\gamma$= 90$^{\circ}$. Fig.~\ref{55} depicts this orthorhombic phase as viewed along the crystallographic $c$-axis. The yellow dotted line indicates the hexagonal arrangement formed by Ir octahedra. The high temperature hexagonal structural symmetry allows for only one crystallographic position (4f) for Iridium. Therefore, given the presence of mixed valent state Ir$^{5.5}$, this position is highly disordered. On the other hand, the  low temperature C-centred orthorhombic symmetry is a 2a x 2b primitive hexagonal pseudo-cell (or an orthohexagonal cell) and allows for three different crystallographic sites for Ir (8f,8f,16h) making it possible for the charge to be redistributed at these distinct cation sites. In addition, Na also now has 3 unique Wyckoff positions  (4a, 4b 8d) allowing for the movement of Iridium while still maintaining the orthorhombic crystal framework. This is a complex low symmetry where each element has multiple unique positions, the details of which are given in  \cite{PhysRevLett.108.217205}. There have been prior reports of orthorhombic phases with only one crystallographic position for Ir, but attempts to fit the low temperature profile of Ba$_{3}$NaIr$_{2}$O$_{9}$ using this symmetry were not successful. Interestingly, in the ruthenium analogue, this need for multiple Ru positions was attributed to the presence of a charge ordered state.\\

\begin{figure}[ht]
	\centering
	\includegraphics[scale=0.54]{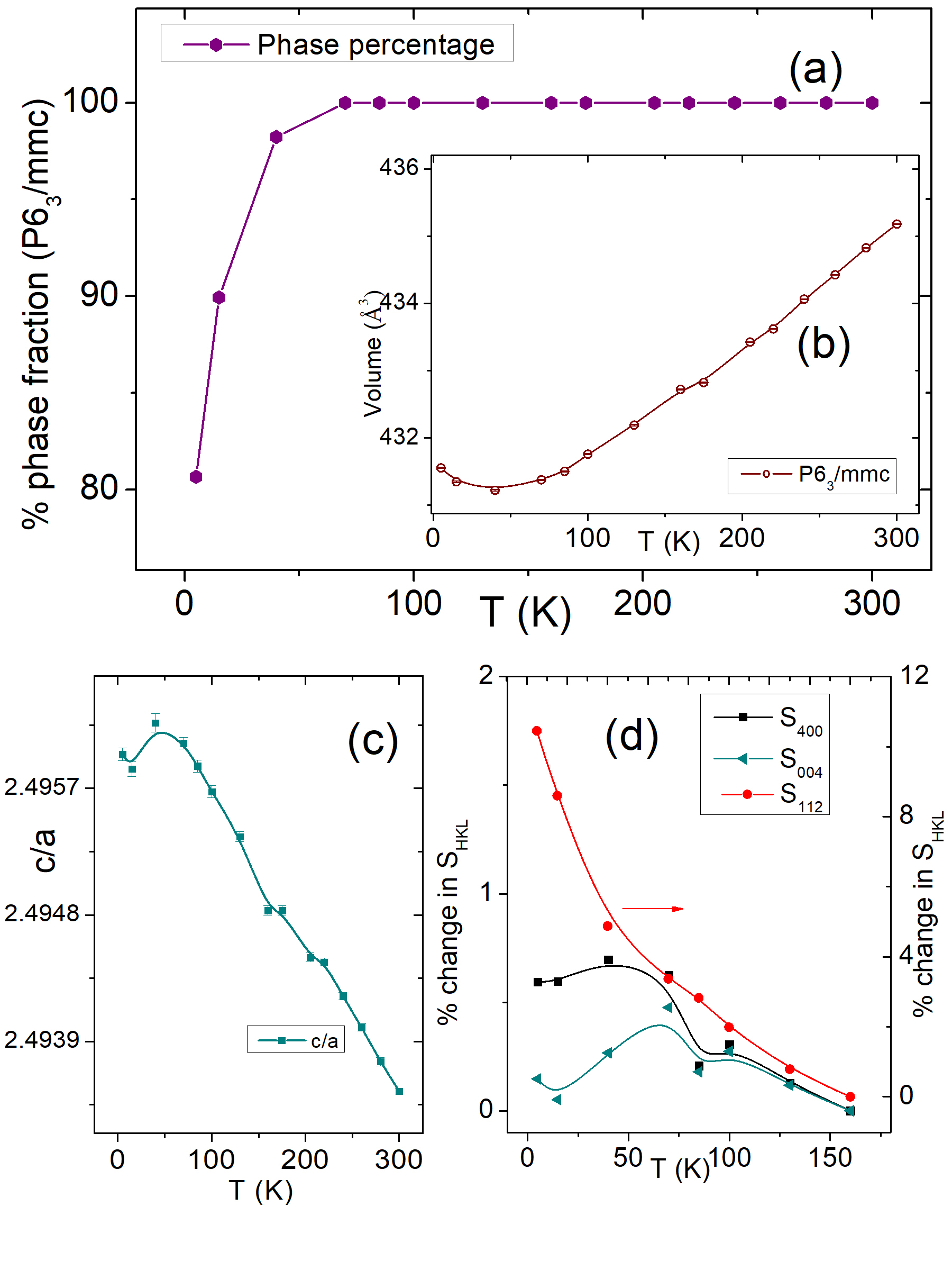}
	\caption{(a) The variation of the phase fraction of the hexagonal P6$_{3}$/mmc with temperature. As the temperature reduces, the hexagonal phase converts slowly to orthorhombic phase, nucleating at 50K and reached 80$\%$ of the total volume fraction at 5\,K.  Temperature evolution of the (b) volume, (c)  ratio of lattice parameters c/a for the hexagonal symmetry. A slight variation in both the parameters are observed marking the onset of the lower symmetry orthorhombic phase. (d) The temperature dependence of the microstrain parameters S${_{HKL}}$ for three different hkl is depicted. The sharp change in S$_{400}$ and S$_{004}$ close to the structural transformation temperature is consistent with distortions of the lattice with the onset of orthorhombic symmetry.}
	\label{56}
\end{figure}
It is observed that down to 50\,K, a single structural hexagonal model with strain parameters is sufficient for the fitting. As a function of reducing temperatures, the phase fraction of hexagonal symmetry is invariant till 50\,K, below which the orthorhombic symmetry is seen to stabilize, reaching 20$\%$ of the total volume fraction at 5\,K (Fig.~\ref{56}(a)). The temperature dependence of volume and $c/a$ ratio for the primary hexagonal phase are depicted in in Fig.~\ref{56}(b) and Fig.~\ref{56}(c) respectively. Clearly, below 50\,K, the $c/a$ ratio shows a change in slope associated with onset of the partial structural phase transformation. The evolution of the secondary orthorhombic phase is also evident in the temperature dependence of the microstrain covariance parameters as in depicted in Fig.~\ref{56}(d).  The strain parameters S$_{400}$ and S$_{004}$ show a sharp change close to the structural transformation temperature and remains almost constant below it, whereas the parameter S$_{112}$ increases dramatically. These changes in the microstrain parameters are indicative of deviations in the $\alpha$ and $\beta$ angles of the hexagonal lattice framework, and consistent with a distortion towards an orthorhombic symmetry. It is interesting to note that the emergence of the secondary orthorhombic phase at low temperatures is not associated with the observation of a splitting of the hexagonal peaks, as was observed in an earlier report on the same system \cite{ZURLOYE2009608}. We believe that this is due to the excess broadening of the diffraction peaks due to strain. This incipient strain not only masks the peak splitting expected due to the orthorhombic distortion, but also results in an incomplete conversion of the high temperature hexagonal phase to the lower symmetry orthorhombic one. \\

\begin{figure}
	\centering
	\includegraphics[scale=0.52]{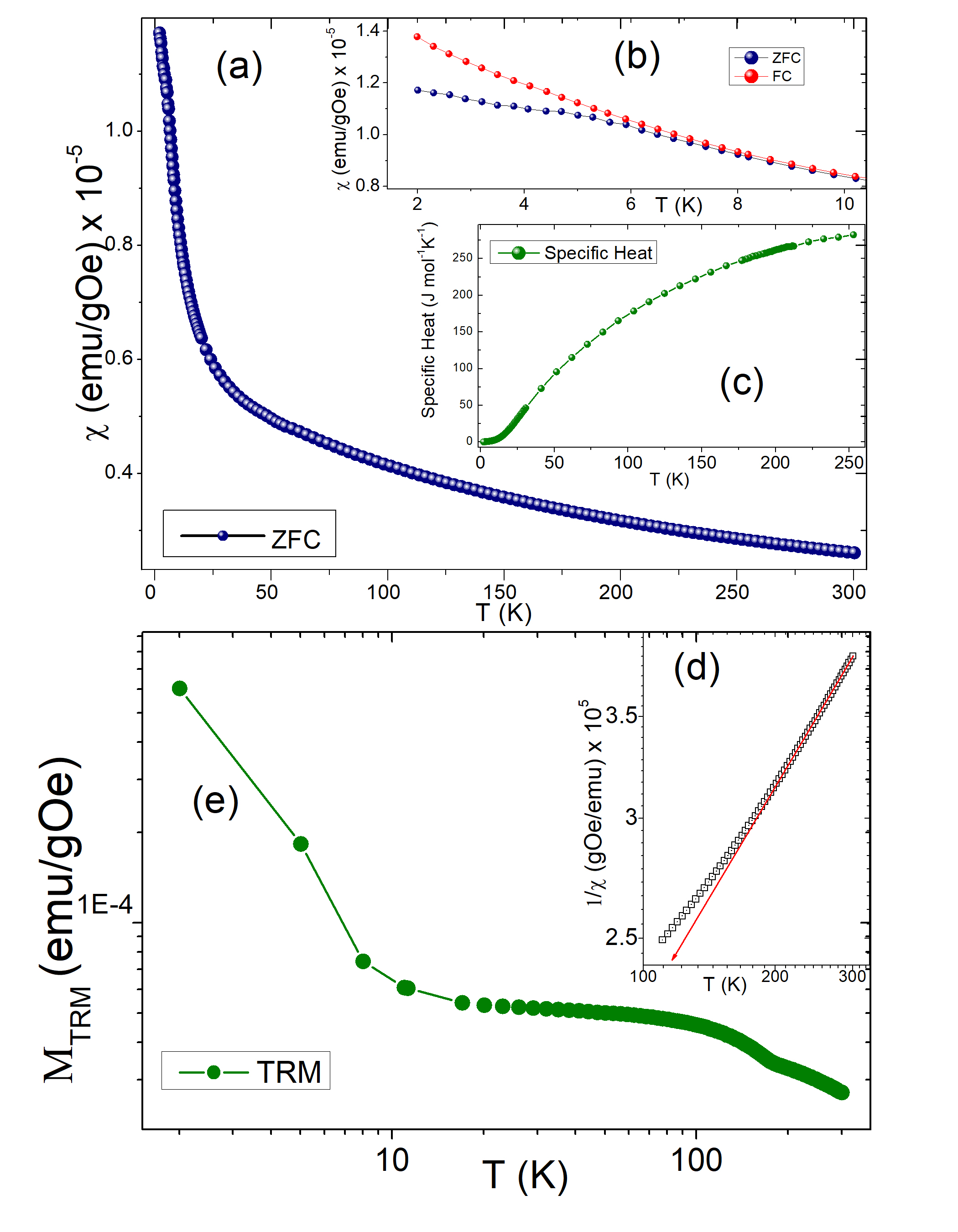}
	\caption{(a) Zero field cooled temperature dependent magnetization measured at 500\,Oe for Ba$_{3}$NaIr$_{2}$O$_{9}$. (b) the FC and ZFC curves show divergence close to 6K, corresponding to a cluster glass transition. (c) Heat capacity as a function of temperature measured in zero magnetic field shows no discernible anomaly in the entire range of measurement. (d) log-log plot of temperature dependence of the inverse magnetic susceptibility data as measured at 500\,Oe. The solid red line is a guide to the eye to show the deviation from Curie-Weiss law, which starts close to 175\,K. (e) Thermo-remnant magnetization (TRM) measured at 1\,kOe with two systematic jumps corresponding to the onset of the co-operative paramagnetic regime, and the cluster glass state respectively.}
	\label{57}
\end{figure}

Fig.~\ref{57}(a) shows the temperature dependence of the magnetic susceptibility of Ba$_{3}$NaIr$_{2}$O$_{9}$ as measured at an applied field of 500\,Oe. The susceptibility increases with decrease in temperature with the zero field cooled (zfc) and field cooled (fc) curves diverging close to 6K as shown in Fig.~\ref{57}(b). This is at variance with what has been reported in early single crystalline specimens of this system, where features in the magnetization was observed at 75\,K and 50\,K \cite{ZURLOYE2009608,Ba3LiNaK}. The temperature dependence of the heat capacity as measured from 2-250\,K is depicted in Fig.~\ref{57}(c). Clearly, the low temperature anomaly observed in magnetization is absent here which implies that the change in entropy is rather small. Fig.~\ref{57}(d) shows the temperature dependence of reciprocal magnetic susceptibility (1/$\chi$). Interestingly, a linear region was observed well in excess of 200\,K, and hence only the temperature range 260- 300\,K was chosen to fit the inverse magnetic susceptibility using the Curie-Weiss law. An effective magnetic moment value 3.42(5)$\mu_B$ per formula unit and a Weiss temperature ($\theta_{c}$) of -285.36(1.1)\,K were obtained,  with the latter being indicative of the extent of frustration in this system, since we only observe a feature in magnetization at 6\,K.  \\

Since Iridium is the only magnetic ion in this system, the magnetic moment arises from the charge balance between Ir(V) (\textit{5d$^{4}$}) and Ir(VI) (\textit{5d$^{3}$}). Based on these oxidation states and the octahedral coordination environments, the theoretical spin-only moment for non-interacting mixed valent Ir$^{5+}$ (S=1, 2.83 $\mu_B$) and Ir$^{6+}$ (S=3/2, 3.87 $\mu_B$) is 6.7$\mu_B$ per formula unit. These calculated moments are significantly larger from the experimentally determined value 3.42(5)$\mu_B$ per formula unit. However, the experimentally obtained value is close to the reported magnetic moments 3.6$\mu_B$ per formula unit for Ba$_{3}$NaIr$_{2}$O$_{9}$ and 3.93$\mu_B$ per formula unit for Ba$_{3}$LiIr$_{2}$O$_{9}$, both having Ir in a similar 5.5 charge state \cite{Ba3LiNaK}. Such reduction in moment is a peculiar feature seen in iridates and has been reported for a wide range of iridium based oxides \cite{PhysRevB.97.064408,Boseggia_2013,PhysRevB.98.245123,annurev-conmatphys-031115-011319}.\\ 

\begin{figure}
	\centering
	\includegraphics[scale=0.4]{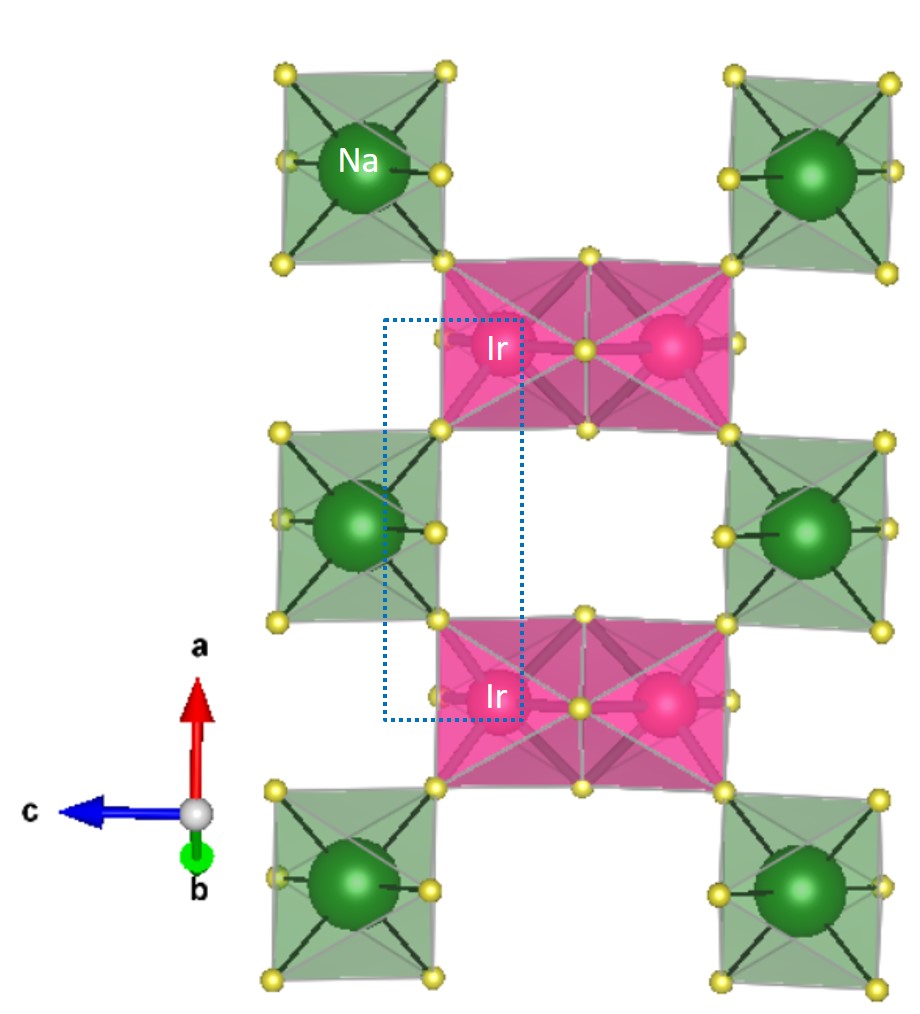}
	\caption{ A schematic representation of the crystal structure of  Ba$_{3}$NaIr$_{2}$O$_{9}$ using Vesta. The projection of the structure perpendicular to the c-axis is shown. Here pink and green octahedra represents Iridium and Sodium respectively and the Barium atoms are not shown for clarity. The Iridium dimers (pink) are separated by Sodium octahedra (green) along the c-axis where the extended super exchange between Ir dimers is mediated by oxygens in Na octahedra as shown by dotted line.}
	\label{58}
\end{figure}

The strong suppression of the magnetic moment here is ascribed to the joint effect of spin orbit interaction and strong covalency, resulting in the formation of metal-metal bonds. They act against the intraatomic Hund's rule exchange interaction to reduce the total magnetic moment on the Iridium dimer. This was further confirmed by our synchrotron measurements where an anomalous shortening of Ir-Ir bond distance in the +5.5 valence state (2.73$\AA$) as compared to the +5 state (2.75$\AA$) corroborates the formation of the metal-metal bonds. Na being non-magnetic, the inter and intra dimer interactions between Ir ions drives the magnetic ground state of the system. In the absence of a superexchange path for inter dimer interactions (J$_{2}$ and J$_{3}$), the extended superexchange pathways Ir-O-O-Ir could possibly influence the magnetic exchange interactions. The Ir dimers are separated from each other via non magnetic Na octahedra (green) as shown in Fig.~\ref{58}. The next nearest neighbour inter dimer Ir (5.8619(8) and 5.6911(12)) are connected via 2 oxygens from the Na octahedra as shown by the dotted lines. Thus, in addition to metal-metal bonds, the  presence of super exchange and extended super exchange interactions pathways lead to complex magnetic exchange interactions.\\
 \\
Though heat capacity measurements did not show evidence of any long range ordered state, a Curie Weiss fit of the inverse magnetic susceptibility was valid only in temperatures in excess of 260\,K, indicating the presence of an extended regime of short range magnetic correlations. To gain further insight in to the extent of this regime, we performed temperature dependent measurements of the Thermo-remnant magnetization (TRM), which has proven to be an effective tool in the investigation of magnetically frustrated systems. A TRM measurement as performed on the Ba$_{3}$NaIr$_{2}$O$_{9}$ system in a cooling field of 1\,kOe is depicted in Fig.~\ref{57}(e). Two precipitous jumps are clearly observed - one below 10\,K, which corresponds to the low temperature magnetic transition observed in the ZFC-FC measurements, and one just below 175\,K, which roughly corresponds to the region where the inverse magnetic susceptibility deviates from the linear Curie-Weiss fit.  In the absence of long range order, this feature at high temperature could be ascribed to the onset of a cooperative paramagnetic regime. First coined by Villain \cite{villain1979insulating}, cooperative paramagnetism was used to describe the low temperature dynamics of a classical Heisenberg spins on a corner sharing tetrahedral framework, and is a defining feature of systems with high geometric frustration. Cooperative paramagnetism is seen in many transition metal oxides which crystallizes in magnetic spin configurations that are geometrically or topologically prone to frustration due to underlying lattices based upon corner, edge or face sharing triangles or tetrahedra. A wide range of systems including pyrochlore, spinels, and jarosites are now known to exhibit this phenomena \cite{doi:10.1143/JPSJ.79.011004,PhysRevB.81.060408,PhysRevB.77.020405}. This state can also be looked upon as being analogous to the Griffiths phase \cite{PhysRevLett.124.046404}, with the notable difference that the low temperature magnetic ground state instead of being magnetically ordered, now undergoes a glass-like dynamical phase transition.  We believe that the nucleation of finite size correlated regions within the antiferromagnetic matrix starts to develop close to 175\,K.  As the temperature reduces, magnetic frustration develops due to competing intra dimer (nearest neighbour J$_{1}$) and inter dimer (next- nearest neighbour J$_{2}$ and J$_{3}$) interactions. The absence of conventional long range antiferromagnetic order is due to the interplay between frustration and quenched disorder. As proposed by Imry and Ma \cite{PhysRevLett.35.1399}, a random quenched disorder inhibits the transition to a long range magnetically ordered state but instead favours the nucleation of correlated magnetic clusters \cite{doi:10.1063/1.5094905}.  \\
\begin{figure}
	\centering
	\includegraphics[scale=0.29]{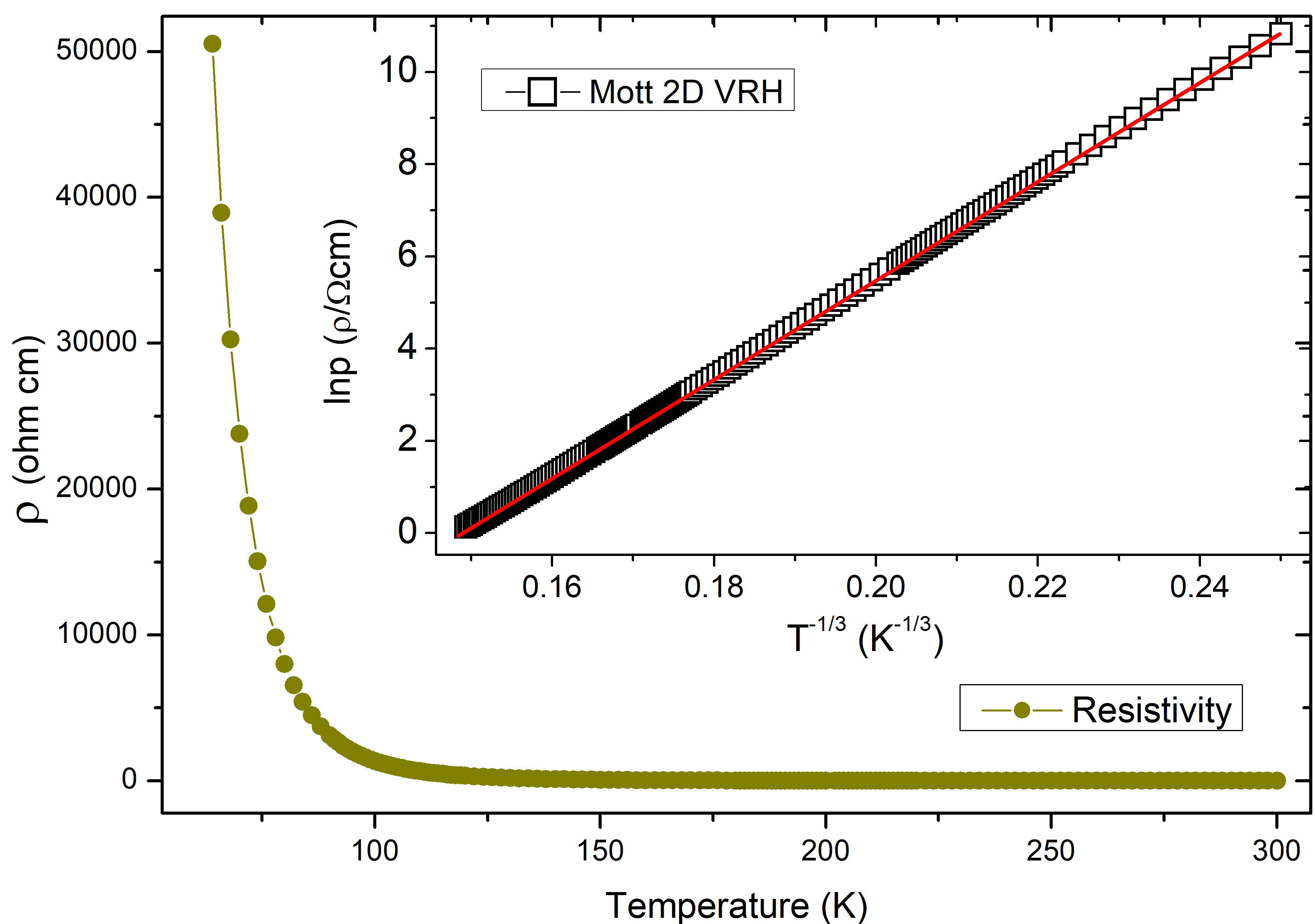}
	\caption{Main panel: Resistivity ($\rho$) plotted as a function of temperature. Inset: Resistivity (ln$\rho$) as a function of temperature (T$^{-0.33}$).  The red line is the fit to the Mott variable range hopping (VRH) model.}
	\label{59}
\end{figure}

Interestingly, the high temperature magnetic feature ($\sim$175\,K) which we observe in the TRM measurements is not easily discernible in other measurements, and hence has gone unreported in prior single crystal measurements of Ba$_{3}$NaIr$_{2}$O$_{9}$  as well \cite{Ba3LiNaK}. This is a consequence of the fact that the magnetic susceptibility of the paramagnetic matrix ($\chi{_{PM}}$) would be of the same order (or even larger) than that of the antiferromagnetic clusters ($\chi{_{AFM}}$), making it difficult to unambiguously determine the contribution of the antiferromagnetic clusters in traditional in-field magnetic measurements. On the other hand, since TRM is a zero-field measurement, the contribution of the paramagnetic magnetic susceptibility is likely to be suppressed, allowing for one to identify more clearly the temperature regime at which the antiferromagnetic clusters begin to nucleate. Interestingly, the ruthenate analogue of this triple perovskite was reported to exhibit the opening of a charge gap at 210\,K \cite{PhysRevLett.108.217205}, though we do not observe any evidence of a similar phenomena in its Iridium counterpart investigated here. The magnetic transition in these family of oxides is typically associated with a symmetry breaking lattice distortion which alters the exchange parameters, thereby neutralizing the frustration. In the case of Ba$_{3}$NaIr$_{2}$O$_{9}$, the interesting capacity of the system to accommodate strain impedes a traditional structural transformation. Therefore, rather than observing a first order transition from hexagonal symmetry to an orthorhombic one, we observe a slowly evolving strained lattice gradually transforming to a lower symmetry where the major phase still retains the original high temperature symmetry.  A strained lattice of this nature is probably closer to the reports of the triple perovskite family when subjected to external pressure \cite{ZHAO2009327,Senn_2013}. For instance on application of pressure, the Ba$_{3}$NaRu$_{2}$O$_{9}$ transforms to a new phase, 3C$_{1:2}$-Ba$_{3}$NaRu$_{2}$O$_{9}$, where the charge gap completely disappears and Pauli paramagnetism emerges, possibly as a consequence of strong electron-electron correlations. The Ba$_{3}$CaRu$_{2}$O$_{9}$ system has also been reported to exhibit excess strain in the lattice, in the form of peak broadening as the temperature was lowered.  Therefore, the ground state in Ba$_{3}$NaIr$_{2}$O$_{9}$  is clearly influenced by the complex interplay of a mixed valent Ir, frustration, phase coexistence and strain.\\

The temperature dependence of the electrical resistivity is shown in the Fig.~\ref{59}. The system is semiconducting in nature, with the magnitude of resistivity changing by 4 orders of magnitude from its room temperature value. Attempts to fit using the Arrhenius model and Efros-Shklovskii variable range hopping (ES-VRH) model were unsuccessful. A better fit was obtained by using the Mott variable-range hopping (VRH) model which is given by:
\begin{equation}
\nonumber
	\rho\   \propto\   exp((T_{0}/T)^{\nu})
\end{equation}
where the best fits were obtained for $\nu$=1/3 indicating variable range hopping in two dimensions \cite{mott2012electronic}. The magnetic Ir dimers are connected via non-magnetic Na octahedra, generating a pseudo 2D structure. Thus, the crystal structure of this triple perovskite can be expressed by alternate stacking of two kinds of 2-D layers which consist of the NaO$_{6}$ octahedra and the Ir$_{2}$O$_{9}$ polyhedra. This may account for the observed 2-D resistivity behaviour. The resistivity of Ba$_{3}$NaIr$_{2}$O$_{9}$ as a function of ln$\rho$ vs T$^{-0.33}$ is shown in the inset of Fig.~\ref{59}. The localization of states due to strong Coulomb interactions and slight structural disorder would be consistent with variable range hopping behaviour. The resistivity of the ruthenate analogues of the triple perovskites Ba$_{3}$MRu$_{2}$O$_{9}$ (R=Fe,Co,Ni,Cu,In,Ln), all follows  the same characteristics \cite{PhysRevB.58.10315,doi:10.1246/bcsj.76.1093}. \\

\begin{figure}
	\centering
	\includegraphics[scale=0.29]{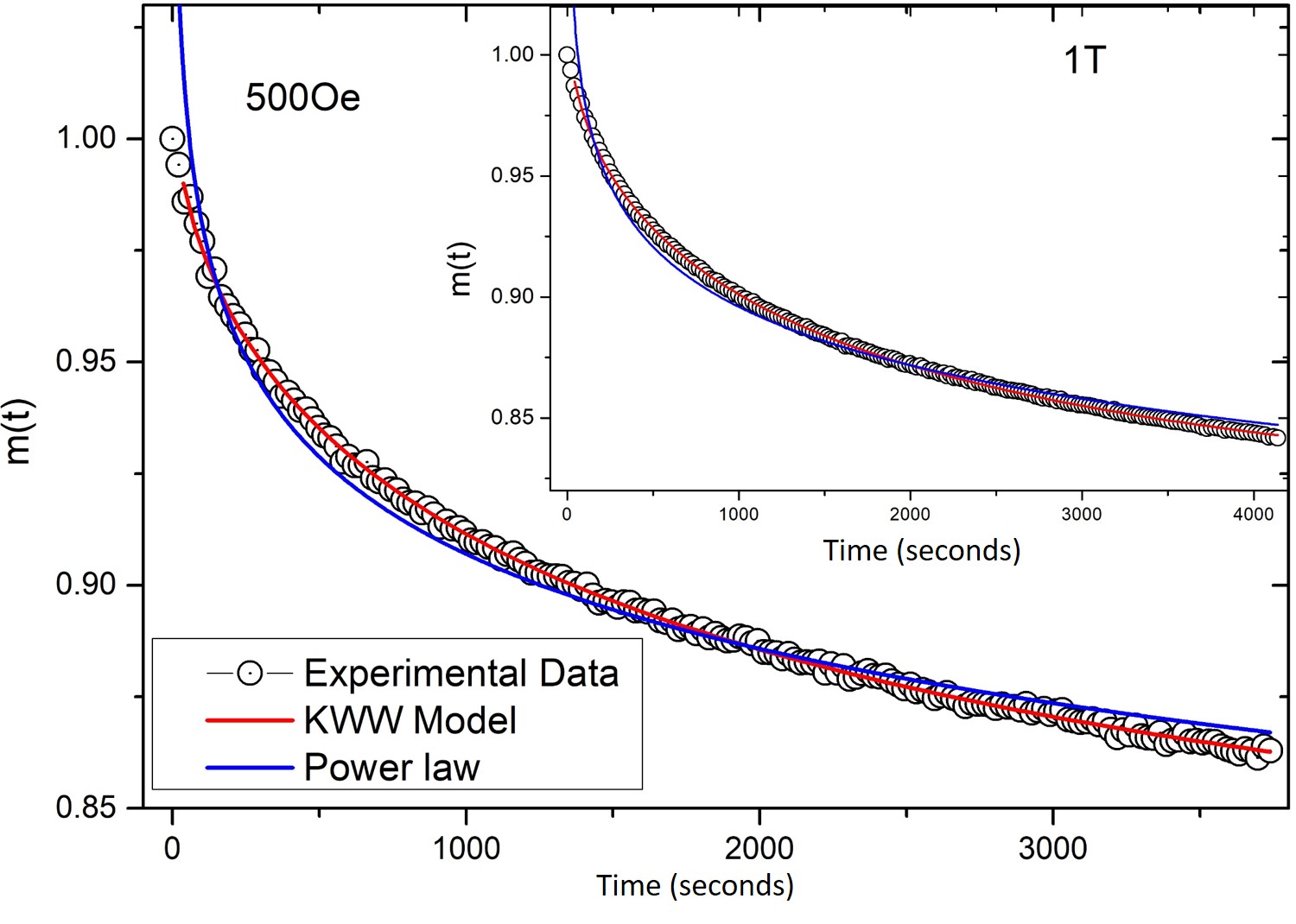}
	\caption{Normalised Isothermal Remanent Magnetization (IRM) at 2K in cooling fields of 500\,Oe and 1\,T (inset), fitted using the Kohlrausch Williams Watt (KWW) stretched exponential (red) and a power law (blue).}
	\label{510}
\end{figure}
\begin{figure}
	\centering
	\includegraphics[scale=0.29]{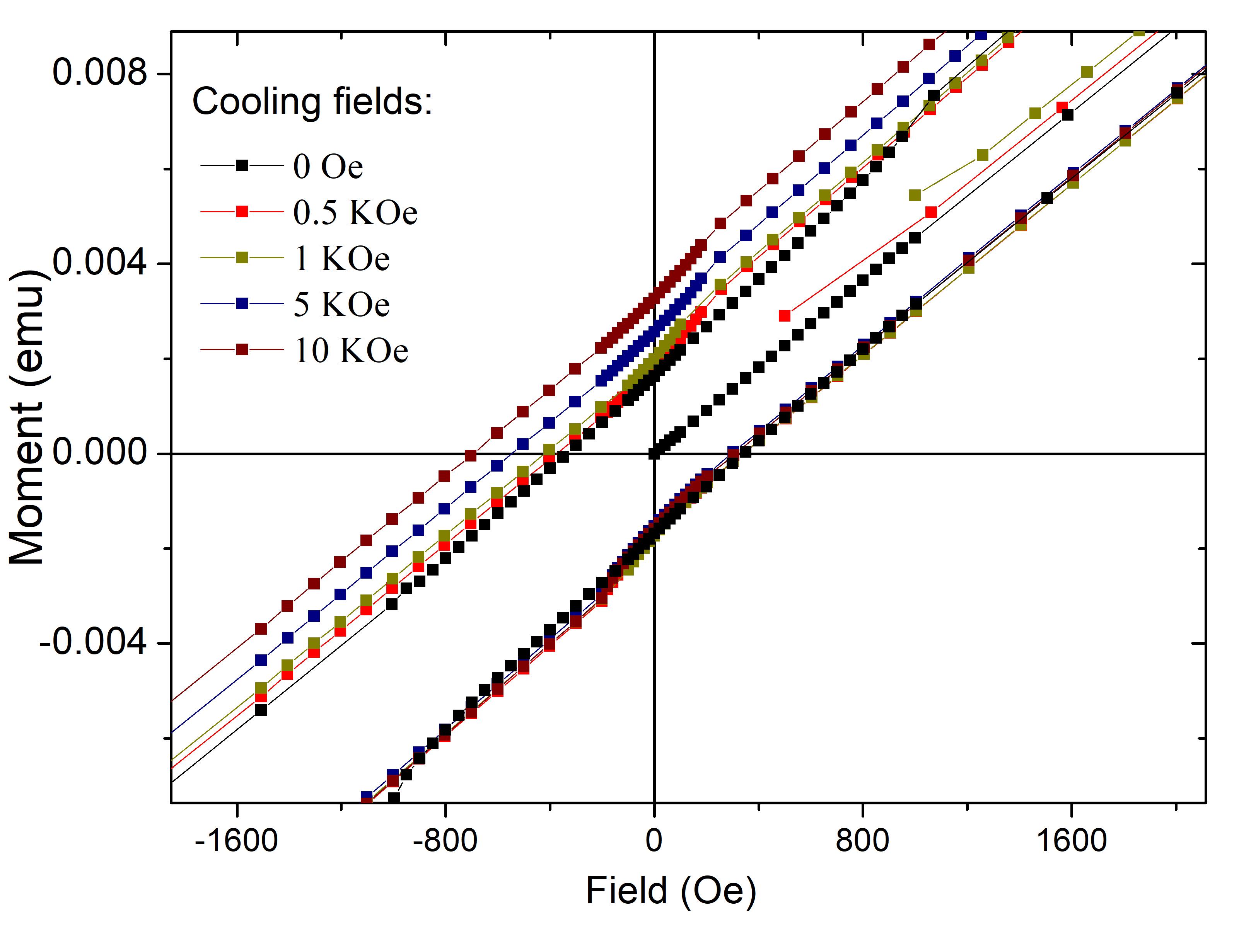}
	\caption{Isothermal magnetization measured at 2\,K for Ba$_{3}$NaIr$_{2}$O$_{9}$ after cooling in different magnetic fields. A systematic shift of the hysteresis loop as a function of the magnitude of the cooling field can be clearly seen indicating the presence of mixed magnetic interactions.}
	\label{511}
\end{figure}
In addition to frustration, the presence of mixed valent states of Iridium and phase co-existence sets the stage for inhomogeneous magnetic exchange interactions. The observation of an anomaly below 6\,K in the M(T) curves indicates the emergence of glassy dynamics owing to this frustration. However, the signal was too small for us to clearly identify a frequency dependent peak in the ac susceptibility measurements. Another method to probe glassy dynamics is to use the time evolution of Isothermal Remanent Magnetization (IRM). This involves cooling the sample from 300\,K to 2\,K in the presence of a magnetic field, after which the magnetic field is switched off and the decay of magnetization is measured as function of time. A special formulation of power law is known to study the time dynamics of magnetization for glasses under stress known as the Kohlrausch Williams Watt (KWW) stretched exponential equation \cite{Edwards_1986,PhysRevLett.57.483,PhysRevB.90.024413} given by :
\begin{equation}
\nonumber
m(t) = m_{0} - m_{g}exp\lbrace-(t/\tau)^{\beta}\rbrace
\end{equation}
where m$_{0}$ is related to initial remanent magnetization, m$_{g}$ is magnetization of glassy component, $\tau$ and $\beta$ are the characteristic relaxation time constant and stretching exponent respectively. Here m(t) is representative of the sum of many exponential decays weighted by a distribution of individual relaxation times, with the magnitude of $\beta$ indicating the breadth of that distribution \cite{SIDEBOTTOM1995151}. The value of $\beta$ has been reported to lie between 0 and 1 for a wide range of disordered systems. The normalized magnetization m(t) = (M$_{t}$/M$_{t=0}$) as measured in Ba$_{3}$NaIr$_{2}$O$_{9}$ at 2\,K with cooling fields of 500\,Oe (main panel) and 1T (inset) at 2K is plotted in Fig.~\ref{510}. As depicted by the blue curve, the fit to a simple power law was not satisfactory. However, a good fit was obtained for the KWW model and the resultant values of $\beta$ are 0.518(14) and 0.5464(68) for 500Oe and 1T respectively. These values are in lines with the reported values for cluster glass phase in many double perovskites \cite{doi:10.1063/1.5094905,anand2019emergence}, and reinforces our contention that the low temperature magnetic ground state is one which has magnetically frozen clusters.\\
\begin{figure}
	\centering
	\includegraphics[scale=0.35]{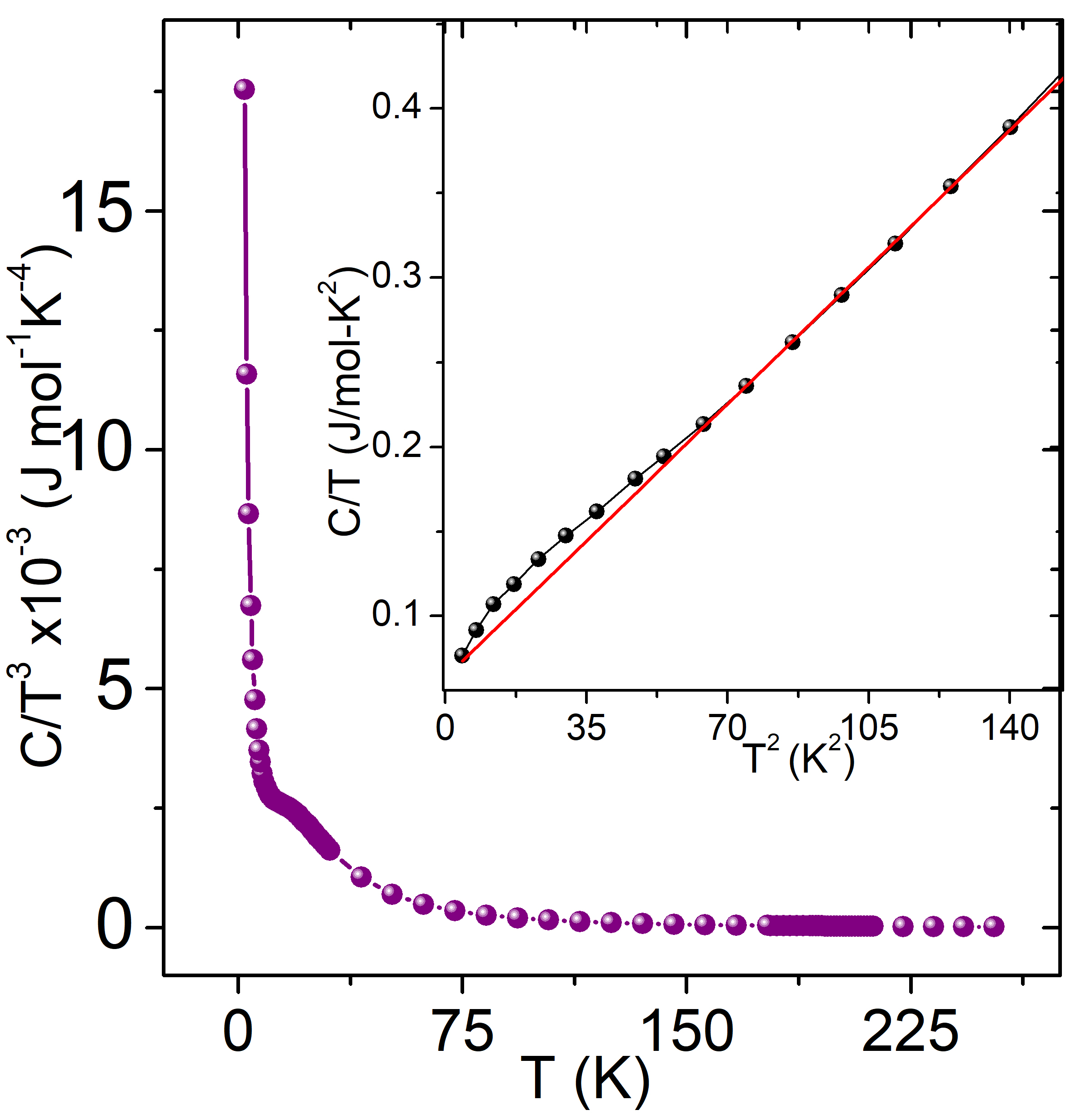}
	\caption{ Main panel: The low temperature specific heat C$_{P}$/T$^{3}$ as a function of temperature. The slight upturn in the low temperature range is a strong indication of disorder in the system.  Inset: The red line line depicts the fir to the low temperature specific heat data using the Debye and Petit's law.  }
	\label{512}
\end{figure}
The magnetic interactions in Ba$_{3}$NaIr$_{2}$O$_{9}$ are predominantly antiferromagnetic, though, signatures of the presence of mixed magnetic interactions are suggested by a weak loop opening in magnetization isotherms at 2\,K as shown in Fig.~\ref{511}.  As revealed by our synchrotron studies, the low temperature structure comprises of a highly strained lattice with two unique structural motifs coexisting. This coupled with the existence of finite size antiferromagnetic clusters allow for exchange bias, with the antiferro and ferro-magnetic contributions arising from the magnetic order within ordered clusters and uncompensated spins at the surface of these clusters respectively (Fig.~\ref{511}). The presence of a low temperature glass-like magnetic ground state is also evidenced in a strong upturn in C/T$^{3}$ vs T Fig.~\ref{512}, with a clear deviation from what is expected from Debye's law. This excess entropy arises as a consequence of the glassy dynamics \cite{Cp}, and appears to be a signature common to structural and magnetic glasses. This is also evident on plotting $C/T$ vs T$^{2}$ curve, indicating the presence of an excess entropy that releases 
as a consequence of short range ordering. The inset of Fig.~\ref{512} shows the fit to the low temperature $C/T$ vs $T^{2}$ curve . The data is fitted using the expression 
$  C_{p} = \gamma T + \beta T^{3} $  
where $\gamma$ and $\beta$ are related to the electronic and vibrational degrees of freedom respectively. We also calculated the Debye temperature $\theta_{D}$, which is derived from the expression, $\theta_{D}$ = (12$\pi^{4}$pR/5$\beta$)$^{1/3}$, where R is the ideal gas constant and p is the number of atoms per formula unit. The calculated value of $\theta_{D}$ is 236.84K. The obtained values of $\gamma$ and $\beta$ are 77mJ/molK$^{2}$ and 2.19mJ/molK$^{4}$T respectively. The high value of $\gamma$, unusual for insulating systems, can be attributed to the inherent disorder which affects the spin, charge and orbital degrees of freedom. This has been previously observed in insulating tripe perovskite iridate (Ba$_{3}$ZnIr$_{2}$O$_{9}$-25.9mJ/molK$^{2}$), and manganites \cite{ PhysRevB.67.024401,PhysRevLett097205}. The high value observed here signifies the excess entropy imparted by the frustration and disorder in this oxide owing to the mixed valence state and stress. Interestingly, on the application of a moderate magnetic field (0.5T), no change in the heat capacity was observed (not shown here), which suggests against the presence of paramagnetic impurity centres. \cite{PMID:21505431,PhysRevB.91.024109}. \\

\section{Summary}
In summary, we report on the structure-property relationship in the mixed valent geometrically frustrated triple perovskite iridate Ba$_{3}$NaIr$_{2}$O$_{9}$.  In contrast to what is expected from purely structural considerations, this system stabilizes in a high symmetry hexagonal symmetry at room temperatures. On reducing the temperature, the lattice prefers to be strained rather than distort to a low symmetry phase, as is the norm in this family of materials. Though a low symmetry orthorhombic phase is finally nucleated below 50\,K, this conversion is only partial and the high symmetry hexagonal structure remains the dominant one down to the lowest measured temperatures. Magnetic measurements indicate an extended co-operative paramagnetic regime, which finally freezes to a cluster glass-like phase at very low temperatures, as is also evidenced from magnetization decay and specific heat data. This makes an interesting addition to the family of triple perovskite iridates which exhibit material sensitive physical properties.
\section{Acknowledgements}
S.N. acknowledges DST India for support through the DST Nanomission Thematic Unit Program, SR/NM/TP-13/2016. C.G. and S.N. thank the Department of Science and Technology, India (SR/NM/Z-07/2015) for financial support and the Jawaharlal Nehru Centre for Advanced Scientific Research (JNCASR) for managing the project. 

\bibliography{NaBib}
\end{document}